# Efficient spin injection through a crystalline $AlO_x$ tunnel barrier prepared by the oxidation of an ultra-thin Al epitaxial layer on GaAs

N. Nishizawa[a)], and H. Munekata

Imaging Science and Engineering Laboratory, Tokyo Institute of Technology

4259-J3-15 Nagatsuta, Midori-ku, Yokohama 226-8503, Japan

**ABSTRACT**

We report that an ultra-thin, post-oxidized aluminum epilayer grown on the AlGaAs surface works as a high-quality tunnel barrier for spin injection from a ferromagnetic metal to a semiconductor. One of the key points of the present oxidation method is the formation of the crystalline $AlO_x$ template layer without oxidizing the AlGaAs region near the Al/AlGaAs interface. The oxidized Al layer is not amorphous but show well-defined single crystalline feature reminiscent of the spinel $\gamma$-$AlO_x$ phase. A spin-LED consisting of an Fe layer, a crystalline $AlO_x$ barrier layer, and an AlGaAs-InGaAs double hetero-structure has exhibited circularly polarized electroluminescence with circular polarization of $P_{EL} \sim 0.145$ at the remnant magnetization state of the Fe layer, indicating the relatively high spin injection efficiency ($\equiv 2P_{EL} / P_{Fe}$) of 0.63.

---
[a)] Electronic mail: nishizawa@isl.titech.ac.jp

**I. INTRODUCTION**

There are two important aspects in the study of tunnel barriers for spin injection between a ferromagnetic metal (FM) and a semiconductor (SC). Firstly, the barrier allows quantum transport of spin-polarized carriers across the FM−SC region, which suppresses the backward flow of un-polarized electrons that is unavoidable in the diffusive transport without the barrier[1-3]. Within the limit of the two current resistor model based on the diffusive transport[2,3], spin polarization of a current, $(I_\uparrow - I_\downarrow)/(I_\uparrow + I_\downarrow)$, is proportional to the ratio of conductivity between SC and FM, $\sigma_{SC}/\sigma_{FM}$, which is as small as $10^{-4}$. To circumvent this problem, there have been many experimental studies on spin injection using various types of tunnel barriers, such as a thin Schottky triangle barrier and oxides barriers represented by thin $AlO_x$, MgO, and $SiO_x$ layers[4]. However, the prominent spin injection endurable for device applications has not been achieved so far. In case of the $AlO_x$ barrier with III-V SC[5-9], the values of spin polarization of carriers in SC has extended for wide range, $P_{spin} = (n_\uparrow - n_\downarrow)/(n_\uparrow + n_\downarrow) \approx 0.05 - 0.4$, even at low temperatures at which D'yakonov-Perel spin relaxation is suppressed in SC. This fact suggests that electronic quality of an oxide barrier still needs to be improved at the stage of preparation and characterization.

Another important aspect of a tunnel barrier is to prevent the formation of interface layers as a consequence of chemical reaction between FM and SC. As for the Fe/GaAs system, for example, slightly different processing temperature would result in the formation of various $Fe_xAs_y$ compounds such as diamagnetic $FeAs_2$, antiferromagnetic FeAs and $Fe_2As$[10] at the interface, all of which would significantly reduce the spin polarization at the FM−SC interface. Therefore, another important role of the barrier is to impede those reactions, as well as diffusion and electromigration of constituent elements between FM and SC.

A $SiO_2$ insulator layer on a Si substrate is generally formed by the high temperature oxidation method. This method is not used for III-V compound semiconductors because group-V elements, such as As and P, have high vapor pressures. Other methods, the low-temperature plasma anodic oxidation[11], metal-organic chemical vapor deposition[12], sputtering[13], electron beam evaporation[14], and atomic layer deposition[15], have been studied with the aim to obtain high quality $AlO_x$ layers, but layers obtained from those methods have suffered from a large numbers of interface traps. Density of the interface traps is $\sim 10^{13}$ $cm^{-2}eV^{-1}$ for the $AlO_x$/GaAs systems[16], which is much higher than that ($\sim 10^9$ $cm^{-2}eV^{-1}$) in the $SiO_2$/Si system. Not only for the conventional charge transport but also for the spin transport, interface states are a matter of serious obstacle because they serve spin-independent intermediate points in the tunneling process[17]. Since the surface dangling-bonds and their oxidized bonds are believed to be the origin of the interface-traps on a GaAs surface[18], the desirable method to form an oxide tunnel barrier should not involve the breaking of chemical bonds on the III-V host surface.

We discuss in this paper a method to prepare a high quality AlO$_x$ barrier layer for injecting spins between Fe and GaAs. The preparation method consists of the formation of a crystalline, template AlO$_x$ surface by the growth of an ultra-thin Al epilayer on a GaAs followed by its post-oxidation, and subsequent deposition of a second ultra-thin Al layer and its post-oxidation. Efficiency of spin injection has been assessed optically through circular polarization of electroluminescence (EL) obtained from a lateral-type, spin light emitting diode (spin-LED)[19], through which we report a significant improvement in spin injection efficiency with the crystalline AlO$_x$ barrier.

## II. EXPERIMENTAL

### A. Preparation and characterization of spin-LED structures

Both LED structures and AlO$_x$ layers were prepared by using a molecular beam epitaxy (MBE) system. Prior to the preparation of AlO$_x$ layer, the LED structure, comprising an AlGaAs/InGaAs double heterostructure (DH), was grown on a *p*-type GaAs (001) substrate at the substrate temperature of $T_s$ = 510 ºC in an MBE growth chamber. The DH consists of, from the top, 300-nm *n*-Al$_{0.1}$Ga$_{0.9}$As (Sn ~ 1×10$^{17}$ cm$^{-3}$) / 15-nm undoped Al$_{0.1}$Ga$_{0.9}$As / 500-nm undoped In$_{0.03}$Ga$_{0.97}$As / 20-nm undoped Al$_{0.2}$Ga$_{0.8}$As / 500-nm *p*-Al$_{0.2}$Ga$_{0.8}$As (Be ~ 1×10$^{18}$ cm$^{-3}$) / 500-nm *p*-GaAs (Be ~ 1×10$^{18}$ cm$^{-3}$) / *p*-GaAs (001). After the preparation of the LED structure, a 5.5-Å thick aluminum epilayer was grown on top of the $n$-$Al_{0.1}Ga_{0.9}As$ surface of the DH structure at $T_s \sim 30$ ºC with reduced arsenic background pressure of $1\times10^{-9}$ Torr or less. The Al/DH sample was then transferred to an MBE loading chamber in which an aluminum oxidation process took place. The oxidation process is detailed in the next section. After the formation of an $AlO_x$ layer of approximately 1-nm thickness, the $AlO_x$/DH sample was taken out into an air atmosphere, and transferred to a separate electron-beam evaporation system in which a 100-nm thick polycrystalline Fe layer, a 5-nm thick Ti protection layer, and a 20-nm thick Au electrode layer were deposited subsequently on the $AlO_x$ layer. The spin-LED structure thus prepared was completed by the aging process of the Fe layer at 230 ºC for 1 hour in a nitrogen gas flow.

Reflecting the in-plane magnetic anisotropy of a polycrystalline Fe layer, spins injected into a spin-LED have their spin axes parallel to the in-plane magnetization vector. To transfer the angular momenta of in-plane spins to light through the radiative recombination in the active region, a thick, bulk-type $In_{0.03}Ga_{0.97}As$ layer was grown instead of quantum wells in which the spin axis of the ground, heavy-hole state points along the growth direction[20]. Naturally, as described in the next paragraph, we measured circular polarization of electroluminescence parallel to the magnetization direction through a cleaved sidewall of the spin-LED.

Spin-LED samples were cleaved along the <110> axis into 1-mm square chips, and they were loaded with great care into an optical cryostat with the (110) sidewall facing a cryostat window. The Fe spin-injection electrodes of the test chips were magnetized at 5 K along the [110] direction by an external magnetic field of $H$ = 5 kOe, prior to the EL experiment at the same temperature. The circular polarization of the EL, $P_{EL} = \{I(\sigma^+) - I(\sigma^-)\}/\{I(\sigma^+) + I(\sigma^-)\}$, was measured by the conventional lock-in technique using a photo-elastic modulator operated at 50 kHz[21]. Here, $I(\sigma^+)$ and $I(\sigma^-)$ are the EL intensity of $\sigma^+$ and $\sigma^-$ components, respectively. Since the ratio of the transition probability in a bulk zincblende crystal is 3 : 1 between the transition via the heavy-hole state and that via the light-hole state, $P_{spin} \approx 2 \times P_{EL}$ in which $P_{spin}$ is the spin polarization of injected electrons that recombine in an active $In_{0.03}Ga_{0.97}As$ layer.

### B. Formation of AlO$_x$ barrier layers

Formation process of a 1-nm thick AlO$_x$ barrier layer on an LED surface consists of four different stages. As stated in the previous section, the first stage was epitaxial growth of a 5.5-Å thick aluminum layer on an $n$-Al$_{0.1}$Ga$_{0.9}$As surface at $T_s$ ~ 30 ºC. The growth rate was 200 nm / h, which was within the optimum range for the epitaxial growth of aluminum on a GaAs (001) surface[22]. We believe that an ultra-thin aluminum epilayer forms Al-As bonds at the Al/GaAs interface and protects the GaAs surface without generating interface traps. The sample was then transferred to an MBE loading chamber in which the sample was exposed to the dry air at the atmospheric pressure for over 10 hours. The oxidation at room temperature would not yield extra kinetic energy that allows migration of Al atoms to break the Al-As bonds and generate dangling bonds at the Al/GaAs interface. We expect that oxygen molecules dissociate and form Al-O bonds preferentially with superficial Al atoms. After this first oxidation, the sample was again transferred back to the growth chamber in which a 2.3-Å Al was deposited at $T_s \sim 30$ °C. This was followed by the second oxidation under the same condition as the first oxidation. We have inferred that oxygen penetrates through the top 4 ~ 6 Å region referring to the early work on the analysis of the aluminum metal surface[23]. Therefore, our treatment would not oxidize the interior Al/GaAs interface. During those four stages, surface condition was studied carefully with in-situ reflection high energy electron diffraction (RHEED) when the sample was in the growth chamber.

Our scenario of the aluminum layer oxidation is consistent with the standard enthalpies of reaction and activation. Let us compare two different oxidation reaction paths which are shown schematically in Fig. 1(a): the first scenario (reaction-A) is the oxidation of Al-As bonds in which an oxygen atom is introduced in between Al and As atoms, whereas the second scenario (reaction B) is the oxidation of superficial Al atoms in which an oxygen atom bridges two Al atoms on the surface. The difference in the sum of dissociation energy between the initial and the final states gives the enthalpy of reaction $\Delta H_r^0$, which can be calculated referring to the bond dissociation energy listed in Table 1[25]. The results are represented in eq. (1) and (2):

reaction A : 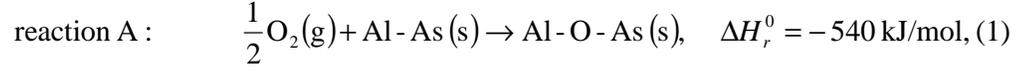 $\frac{1}{2}O_2(g) + \text{Al-As}(s) \rightarrow \text{Al-O-As}(s), \quad \Delta H_r^0 = -540 \text{ kJ/mol}, \quad (1)$

reaction B : 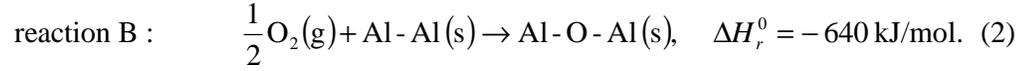 $\frac{1}{2}O_2(g) + \text{Al-Al}(s) \rightarrow \text{Al-O-Al}(s), \quad \Delta H_r^0 = -640 \text{ kJ/mol}. \quad (2)$

Here, influence of the second nearest neighbors and the surface energy are neglect for simplicity. The enthalpy of activation can be regarded as the sum of dissociation energy of the initial state, which can be estimated to be 452 and 382 kJ/mol for the reaction A and B, respectively. As the whole, the reaction B would be the preferred scenario in view of enthalpies of reaction and activation, as shown schematically in Fig. 1(b).

Table I. A list of bond dissociation energy at 298K [25].

| Molecule | Bond dissociation energy (kJ/mol) |
|---|---|
| Al–As | 202.9 |
| Al–Al | 133.0 |
| Al–O | 511.0 |
| As–O | 481.0 |
| O=O | 498.3 |

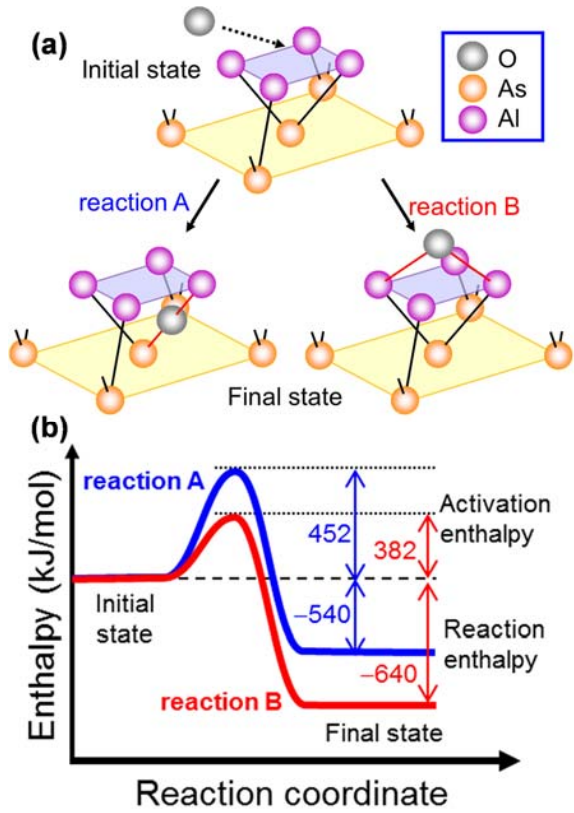

Figure 1: (Color online) (a) Schematic illustrations of two different oxidation reactions occurring around an Al monolayer (ML) chemisorbed on the GaAs interface. The atomic configuration of 1-ML Al on top of the As-terminated GaAs(001) surface is shown schematically in the upper panel. Blue and yellow planes are those representing the Al and As planes, respectively. The oxidation of an Al-As bond (reaction A) is shown in the lower, left panel, whereas the oxidation of an Al-Al bond (reaction B) in the lower, right panel. (b) Schematic illustration of static enthalpy diagram for reactions A and B at room temperature. The values of activation and reaction enthalpies are calculated on the basis of the bond dissociation energy listed in Table 1.

## III. RESULTS AND DISCUSSION

Figures 2 (a) − (e) are pictures of RHEED patterns observed at $T_s \sim 30$ ºC with background pressure of $1 \times 10^{-9}$ Torr or less after completing five different processes; (a) epitaxial growth of a top $n$-Al$_{0.1}$Ga$_{0.9}$As layer, (b) epitaxial growth of a first 5.5-Å thick Al layer, (c) oxidation of the first Al layer, (d) deposition of a second 2.3-Å thick Al layer, and

(e) oxidation of the second Al layer. Pictures in the left column were taken along the GaAs [1$\bar{1}$0] azimuth, whereas those in the right column along the GaAs [110] azimuth,. The pattern shown in Fig. 2(a) is characteristic of an As-stabilized $c(4 \times 4)$ surface. The streaky $(1 \times 1)$ pattern shown in Fig. 2(b) is consistent with those found earlier[22, 23] in sense that an fcc Al layer grows on a GaAs (001) surface with in-plane, 45deg. rotation; namely, $a_{\mathrm{Al_{0.1}Ga_{0.9}As}}^{\mathrm{zincblende}}(110) // a_{\mathrm{Al}}^{\mathrm{fcc}}(100)$: the lattice mismatch between the two planes is about 1.3%, referring to the values 0.7995 nm and 0.8098 nm for $a_{\mathrm{Al_{0.1}Ga_{0.9}As}}^{\mathrm{zincblende}}(110)$ and $2 \times a_{\mathrm{Al}}^{\mathrm{fcc}}(100)$, respectively. To our surprise, the streaky RHEED pattern has lasted throughout the entire oxidation process as shown in Figs. 2(c) – (e), which indicates that the resultant $AlO_x$ layer possesses the crystalline feature with an atomically flat surface.

An interval between streaks in the RHEED patterns shown in Figs. 2(a) to (e) were converted into the spacing of lattice planes parallel to GaAs (1$\bar{1}$0) plane, and are shown in Fig. 2(f) in accordance with the progress of the oxidation process. The spacing of a GaAs (1$\bar{1}$0) plane was used for the internal calibration. The lattice spacing in the first Al layer is nearly the same as that of the $a_{\mathrm{Al_{0.1}Ga_{0.9}As}}^{\mathrm{zincblende}}(110)$ plane, which is consistent with the earlier work reporting the 45-deg. rotation of the fcc-Al lattice to accommodate the condition $a_{\mathrm{Al_{0.1}Ga_{0.9}As}}^{\mathrm{zincblende}}(110) \approx 2 \times a_{\mathrm{Al}}^{\mathrm{fcc}}(100)$[22, 23]. A model of atomic stacks across the Al-GaAs interface is illustrated schematically in the left panel of Fig. 2(g) on the basis of the observed RHEED patterns. We infer that around 70% of aluminum atoms in the nominal 5.5-Å thick Al epilayer are used to completely cover the GaAs(001) surface whereas the remaining aluminum atoms are used for the formation of two-dimensional islands.

The lattice spacing decreases to around 0.67 nm after the first oxidation of the Al epilayer. This value is close to those of the hexagonal $\alpha$-Al$_2$O$_3$ ($11\bar{2}0$) or the spinel $\gamma$-Al$_2$O$_3$ (222); $a^{hcp}_{\alpha\text{-Al}_2\text{O}_3}(11\bar{2}0) = 0.673$ nm and $3 \times a^{spinel}_{\gamma\text{-Al}_2\text{O}_3}(222) = 0.683$ nm, respectively. Since the oxidation was carried out at $T_s \sim 30$ °C, we infer that the surface structure of the oxidized Al epilayer is close to that of the (222) plane of $\gamma$-Al$_2$O$_3$ which is the lower-temperature phase. As shown schematically in the right panel of Fig. 2(g), the oxygen atom, which is chemically adsorbed on the Al surface at room temperature, is stabilized in the form of the Al-O-Al bond which consequently reduces the spatial distance between two nearest neighbor Al atoms through the $s$-$p$ hybridization. That the value of lattice spacing remains close to that of the $3 \times a^{spinel}_{\gamma\text{-Al}_2\text{O}_3}(222)$ after the deposition of the second 2.3-Å thick Al layer suggests that additional Al atoms are accommodated on the surface in the form of Al-O bonds. Knowing that the thickness of naturally oxidized layer on the surface of an Al single crystal is 4 ~ 6 Å[24], we infer that the interior Al/GaAs interface would hardly be oxidized in the present oxidation process.

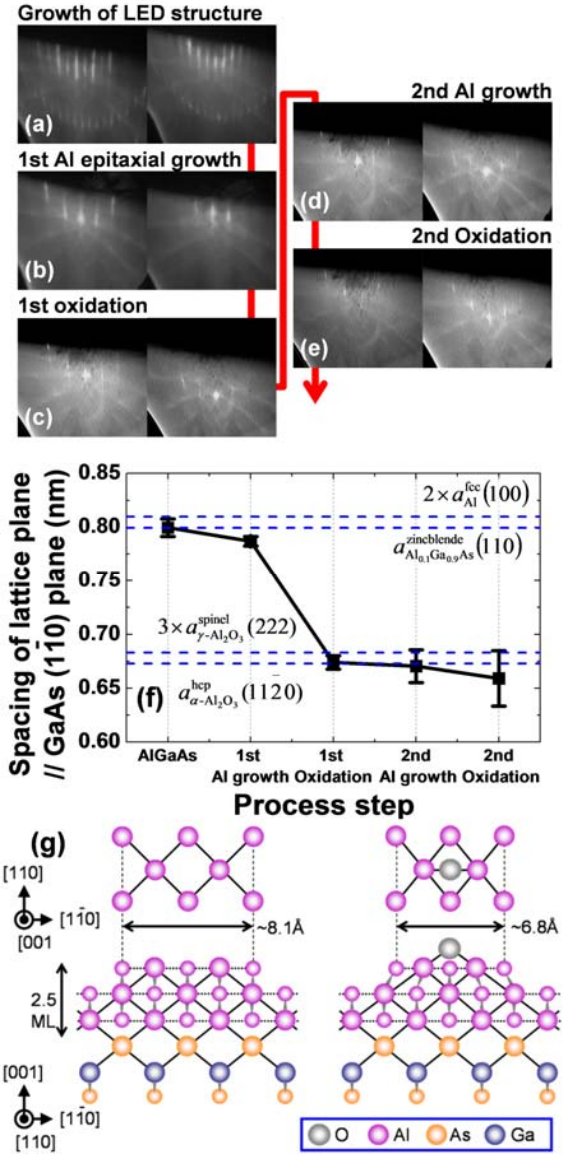

Figure 2: (Color online) RHEED patterns observed along two orthogonal azimuths, GaAs[1$\bar{1}$0] and [110], after completing five different processes; (a) epitaxial growth of a top *n*-AlGaAs layer in a DH, (b) epitaxial growth of a first Al layer, (c) post-oxidation of the first Al epilayer, (d) deposition of a second Al layer, and (e) post-oxidation of the second Al layer. The acceleration voltage of an electron beam was 15 keV. (f) A plot of spacing of lattice planes in accordance with the progress of oxidation process. (g) Schematic illustrations of atomic configurations of (left panels) an Al epilayer and (right panels) an oxidized Al epilayer from the top and side views, respectively. Atoms and bonds aligned on the paper plane are shown by larger circles and solid lines, respectively. Those on the plane a half-ML away from the paper plane are represented by smaller circles and broken lines, respectively.

Crystallographic evaluation around the Fe/AlO$_x$/$n$-Al$_{0.1}$Ga$_{0.9}$As interface in the spin-LED was carried out by using a transmission electron microscope (TEM). A high-resolution cross-sectional view around the interface is shown in Fig. 3(a). We are able to see an ultra-thin, bright region in between a single-crystalline $n$-Al$_{0.1}$Ga$_{0.9}$As and a poly-crystalline Fe layer, with lattice images being different from those in the $n$-Al$_{0.1}$Ga$_{0.9}$As layer. We infer that this bright region is a trace of the oxidized Al epilayer; namely, the crystalline, ultra-thin AlO$_x$ layer. A magnified lattice image near the interface is shown in Fig. 3(b), together with the results of image simulation for $\gamma$-Al$_2$O$_3$(111)/GaAs(110) and fcc-Al(100)/GaAs(110) planes obtained by the calculation with the multi-slice simulation code QSTEM[26] in Figs. 3(c) and (d), respectively. It is clear that a triangular-lattice image in the bright region does not match with the calculated, four-fold lattice image of fcc-Al(100), but is rather close to the calculated, three-fold lattice image of $\gamma$-Al$_2$O$_3$(111). This fact can further be confirmed by locally comparing experimental and calculated images, as shown in the upper panels of Fig. 3(e). Experimental and calculated images shown in the lower panels of Fig. 3(e) for the $n$-AlGaAs region confirm the validity of our simulation in sense that the dumbbell feature consisting of a pair of Ga and As atoms is well reproduced by the simulation. An irregular lattice image right at the AlO$_x$/AlGaAs interface may be viewed as the atomistic transition region consisting of the mixture of $\gamma$-Al$_2$O$_3$ and residual fcc-Al epitaxial domains. With results shown in Figs. 2 and 3, we are able to conclude that oxidized

Al epilayer results in, at least partially, a crystalline AlO$_x$ layer. We hereafter describe this layer as the x-AlO$_x$ layer.

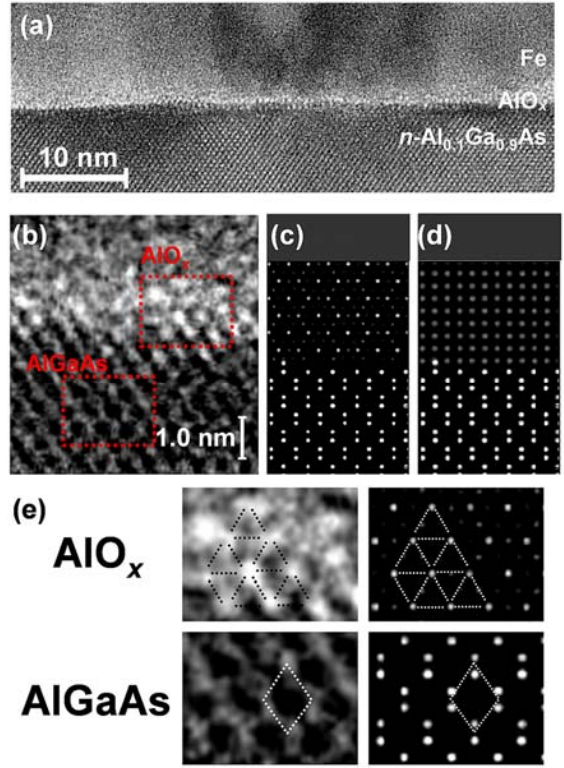

Figure 3: (Color online) (a) Cross-sectional, bright-field TEM image around Fe/AlO$_x$/AlGaAs region in a spin-LED structure. Hitachi H-9000NAR was used with an acceleration voltage of 200 keV. Specimens with the cross section of the (110) plane were prepared by the micro-sampling method using a Ga focus-ion beam equipment. (b) Magnified view of the image shown in (a). Calculated lattice images (c) for the $\gamma$-Al$_2$O$_3$(111)/AlGaAs(110) and (d) for the fcc-Al(100)/AlGaAs(110) structures. Calculation was carried out by using simulation code QSTEM[26]. (e) Further magnified views of (upper two panels) the oxidized Al region together with a calculated lattice image for $\gamma$-Al$_2$O$_3$(111), and (lower two panels) the AlGaAs epitaxial region together with a calculated lattice image for AlGaAs(110). Dotted lines are eye guides for visual inspection.

Spatial distribution of Fe, Al, O, Ga and As around the Fe/x-AlO$_x$/AlGaAs interface was analyzed by the scanning electron energy loss spectroscopy (EELS) with an electron beam (200 keV) of 0.2-nm in diameter. The spatial resolution was around 0.4-nm. Relatively high energy-loss threshold energies were used for quantitative analysis: they were 708 eV (Fe-$L_3$), 1563 eV (Al-K), 532 eV (O-K), 1115 eV (Ga-$L_3$), and 1323 eV (As-$L_{23}$), for Fe, Al, O, Ga and As, respectively. The $L_{23}$-shell threshold energy (73.1 eV) of Al was not used because it was close to the $M_{23}$-shell threshold of Fe (54.0 eV). The intensity of EELS signal $I_n$ for each element $n$ was calculated by integrating the electron counts above the threshold energy excluding the exponential-type background[27]. Each $I_n$ value is then converted into the local composition with $I_n / \sum_n I_n$, and is plotted in Fig.4 (a) as a function of distance $d$ with respect to the origin in the AlGaAs region. Here, the relative error of the local compositions is within 10 %. The corresponding TEM image is shown in the upper part of the figure, in which the position of the crystalline $AlO_x$ layer, the bright region, is supposed to be at $d$ = 17 ~ 18.2 nm. In the bulk $n$-AlGaAs region ($d \leq 15$ nm), the EELS signals of gallium, arsenic, and aluminum are nearly constant. As the electron beam approaches the AlGaAs/$AlO_x$ interface ($d$ = 17 nm), the gallium and arsenic EELS signals both decrease abruptly, whereas the aluminum EELS signal extends up to $d$ ~ 19 nm. The iron EELS signal starts increasing at $d$ ~ 15 nm and saturates at $d$ ~ 22 nm. The oxygen EELS signal starts rising at $d$ ~ 13 nm, forms a peak at $d$ ~ 18 nm, and decays as far as $d$ ~ 25 nm beyond which it stays at a small, constant value. While those profiles are roughly consistent with the TEM images, they reveal the presence of iron in $AlO_x$ and AlGaAs layers, and oxygen in the Fe and AlGaAs layers.

The observed small out-diffusion of iron and oxygen may be attributed to the aging process of the Fe layer (230 °C for 1hr), which we hope to suppress by optimizing the deposition process in the future. The small oxygen signals in the bulk Fe region are presumably due to natural oxidation when a TEM specimen was taken out in the air atmosphere.

Figure 4 (b) shows the EELS spectra around the Fe-$L_3$ energy-loss threshold energy at different $d$ values. Exponential-type backgrounds are removed from the spectra. The spectra do not exhibit any chemical shift, and are identical to that of a metal Fe. This fact suggests two important points; firstly, the amount of oxygen in the Fe layer is not large enough to cause the chemical shift, and secondly, the amount of $Fe_xAs_y$ compounds is negligibly small in the AlGaAs region near the interface.

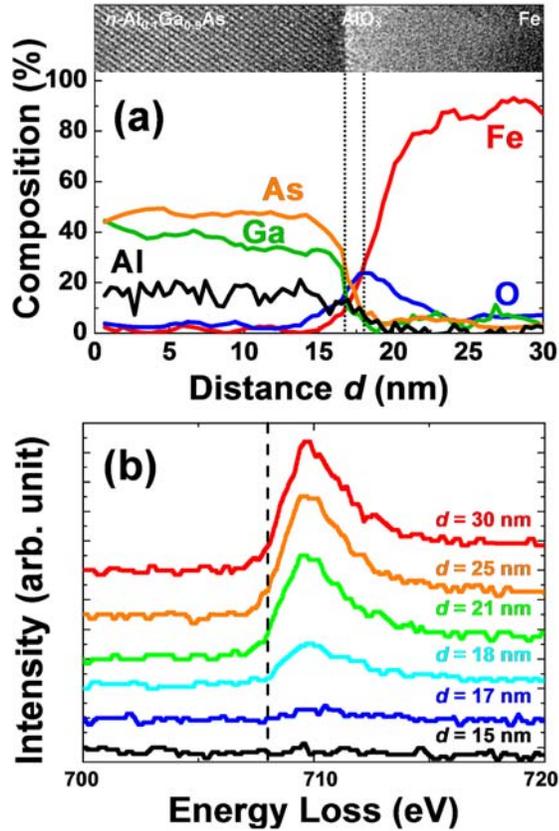

Figure 4: (Color online) (a) Spatial profiles of EELS signals for iron (Fe), aluminum (Al), oxygen (O), gallium (Ga), and arsenic (As) across a Fe/AlO$_x$/AlGaAs interface. The distance $d$ is defined as the position measured from the origin in the AlGaAs region. The corresponding TEM image is shown in the upper part of the figure, in which the position of an AlO$_x$ layer, the bright region, is supposed to be at $d =$ 17 ~ 18.2 nm. (b) EELS spectra around the Fe-L$_3$ threshold energy (708 eV) obtained at different $d$ values.

Carrier transport across the x-AlO$_x$ layer was studied by measuring the $I$-$V$ characteristics of a diode (the diode A) consisting of 100-nm Al / 1-nm x-AlO$_x$ / 300-nm $n$-GaAs:Sn ($N_d$ ~ $1 \times 10^{17}$ cm$^{-3}$) / $n$-GaAs(001). The $I$-$V$ characteristics of another diode (the diode B) consisting of 100-nm Al / 15-nm $n^+$-GaAs (Sn ~ $5 \times 10^{18}$ cm$^{-3}$) / 300-nm $n$-GaAs (Sn ~ $1 \times 10^{17}$ cm$^{-3}$) / $n$-GaAs(001) was also studied for comparison. The triangular Schottky barrier of the $n^+$-GaAs is a tunnel barrier in the diode B. Shown in Fig. 5 are the $I$-$V$ characteristics obtained at 80 K for both diodes. Polarity of the bias voltage is defined with respect to the *n*-GaAs. The diode B shows rectification characteristics, whereas the diode A exhibits almost symmetric characteristics with a current density which is two orders of magnitude less than that of the diode B. The features observed in the diode A strongly suggest that the carrier transport in the diode A is dominated by the tunneling through the x-AlO$_x$ layer. A fit to the positive bias region of the *I-V* curve of the diode A was carried out with the Simmons' equation[28], through which we obtain the average barrier height of 2.8 eV together with the average barrier thickness of 1.0 nm. The value of the barrier height is close to that obtained for a high-quality Al/AlO$_x$ junction prepared by the atomic layer deposition[29], which is very encouraging.

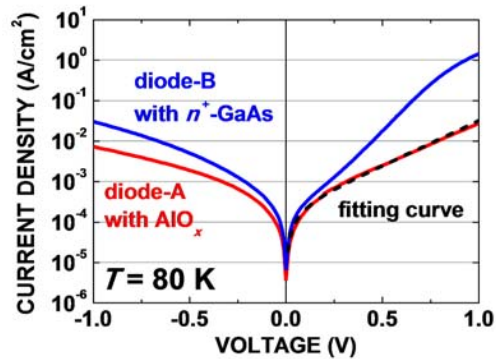

Figure 5: (Color online) *I-V* characteristics of diode A (red lines) and diode B (blue lines) measured at 80 K. A broken line represents the *I-V* curve obtained from the model calculation based on the using Simmons' equation[28].

Helicity dependent electroluminescence (EL) spectra obtained at 5 K from the spin-LED incorporating the Fe/x-AlO$_x$ spin injector (a spin-LED C) is shown in Fig. 6(a).

The spectra obtained from the spin-LED without the $AlO_x$ layer (a spin-LED D) are shown in Fig. 6(b) for comparison. The experimental configuration is depicted in the insets of both figures. For both experiments, spin-injecting Fe electrodes were magnetized along the in-plane easy axis parallel to the GaAs [110] with a magnetic field of $H = 5$ kOe, and no external field was applied during the EL measurements. EL spectra of the spin-LED C show noticeable difference in their peak intensity between right- and left-circular polarizations, whereas those of the spin-LED D exhibit no difference in the peak intensity between the two helicity. The integrated value of circular polarization over the entire EL spectral region is $P_{EL}$ ~ 0.145 for the spin-LED C, which corresponds to $P_{spin}$ ~ 0.29. Knowing that the value of spin polarization has been $P_{Fe}$ ~ 0.46 for a pure Fe layer[30], the overall spin injection efficiency, which we express with the ratio $\varepsilon = P_{spin} / P_{Fe}$, is estimated to be $\varepsilon = 0.63$. The $\varepsilon$ values in spin-LEDs using Fe-based ferromagnets with either an $AlO_x$ barrier or a triangular Shottky barrier have extended for wide range of values at low temperatures[5-9, 31, 32]. The highest $\varepsilon$ value for the spin-LED with $AlO_x$ tunnel barrier has been $\varepsilon = 0.87$ with an external magnetic field of $H = 20$ kOe[9]. The $\varepsilon$ value is affected not only by the spin polarization of a ferromagnet but also by the spin relaxation in semiconductor layers, and can not be compared directly among different device structures. Nevertheless, the $\varepsilon$ value obtained from the spin-LED C has been the highest as far as the spin-LED at the remnant state is concerned.

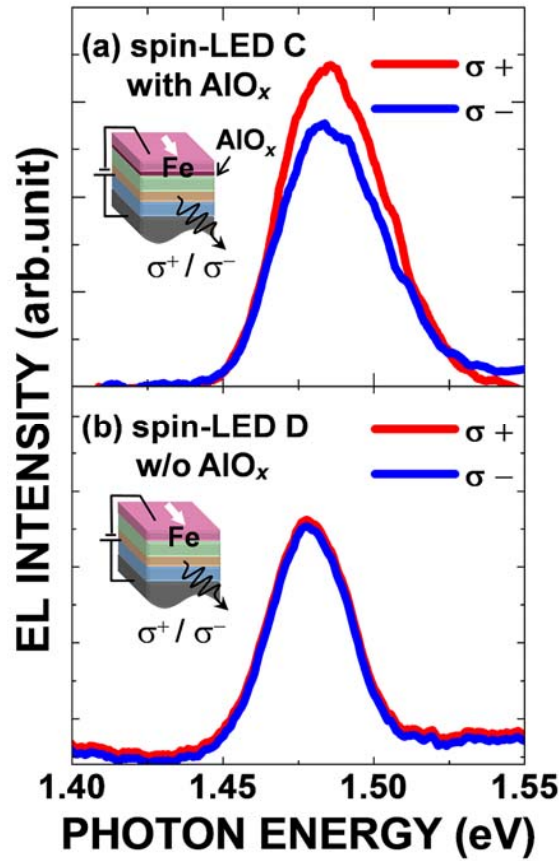

Figure 6: (Color online) (a) $\sigma^+$ and $\sigma^-$ components of EL spectra obtained from (a) spin-LED C and (b) spin-LED D. No external magnetic field was applied during the measurements. Insets are schematic illustrations of experimental configurations showing EL emission from a cleaved side wall and parallel relation between a magnetization vector and an optical axis.

We note here the studies of an aluminum-based, crystalline oxide barrier in other sub-fields. There was a report in the field of superconductor research that a single-crystalline $\alpha$-Al$_2$O$_3$ layer was obtained by the high-temperature oxidation of an amorphous AlO$_x$ layer deposited on the epitaxial superconductor surface, with which a tunnel junction with low current leakage was realized[33]. More recently, it was reported in the field of spintronics research that a tunnel magneto-resistance device with a single crystalline Al$_2$MgO$_4$ barrier exhibited the TMR ration of 188 % at room temperature[34, 35]. Those results, added with the results reported here, have showed importance of studying the preparation and characterization of single-crystalline oxide tunnel barriers for devices utilizing quantum transport.

**IV. CONCLUSION**

Aiming at improving the efficiency of electrical spin injection into GaAs-based structures, a preparation method for the formation of an AlO$_x$ tunnel barrier on a GaAs-based spin-LED has been studied. The strong point of the developed method is gentle, post-oxidation of an Al epilayer grown on top of the surface of III-V compound semiconductors. Crystallographic evaluations with in-situ RHEED and cross-sectional TEM have revealed that an oxidized Al epilayer results in, at least partially, a crystalline AlO$_x$ layer (x-AlO$_x$) whose structure is close to the low-temperature phase $\gamma$-Al$_2$O$_3$. Scanning EELS has been carried out across the Fe/x-AlO$_x$/AlGaAs interface region to evaluate spatial distribution of constituent elements, which has revealed the presence of a thin aluminum-oxide region at the expected location but with a little out-diffusion of oxygen and iron into neighbor layers. However, formation of FeO$_x$ or Fe$_x$As$_y$ interface layers has hardly been observed. A spin-LED consisting of Fe/x-AlO$_x$/DH has exhibited circularly polarized electroluminescence with circular polarization of $P_{EL}$ ~ 0.145 ($P_{spin}$ ~ 0.29), which indicates the highest spin polarization among spin-LEDs having Fe-AlO$_x$ combination. With these experimental results, we have concluded that a high-quality tunnel barrier for spin injection into a GaAs-based structure has been realized with the oxidized Al epilayer. Using other ferromagnetic metals of higher spin polarization together with the crystalline-AlO$_x$ barrier would further improve the performance of spin-LEDs in terms of circular polarization of electroluminescence.

**ACKNOWLEDGMENTS**

Authors are grateful to T. Takita and N. Hieda for their technical assistance in various experiments. We acknowledge partial supports from Advanced Photon Science Alliance Project from MEXT and Grant-in-Aid for Scientific Research (No. 22226002) from JSPS.

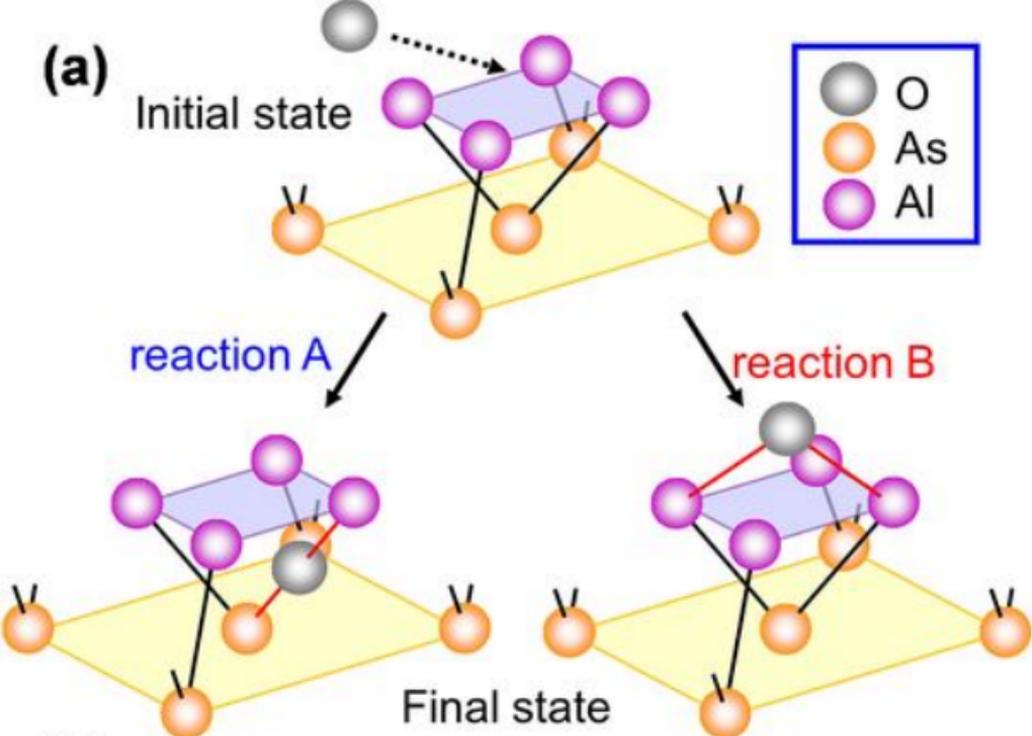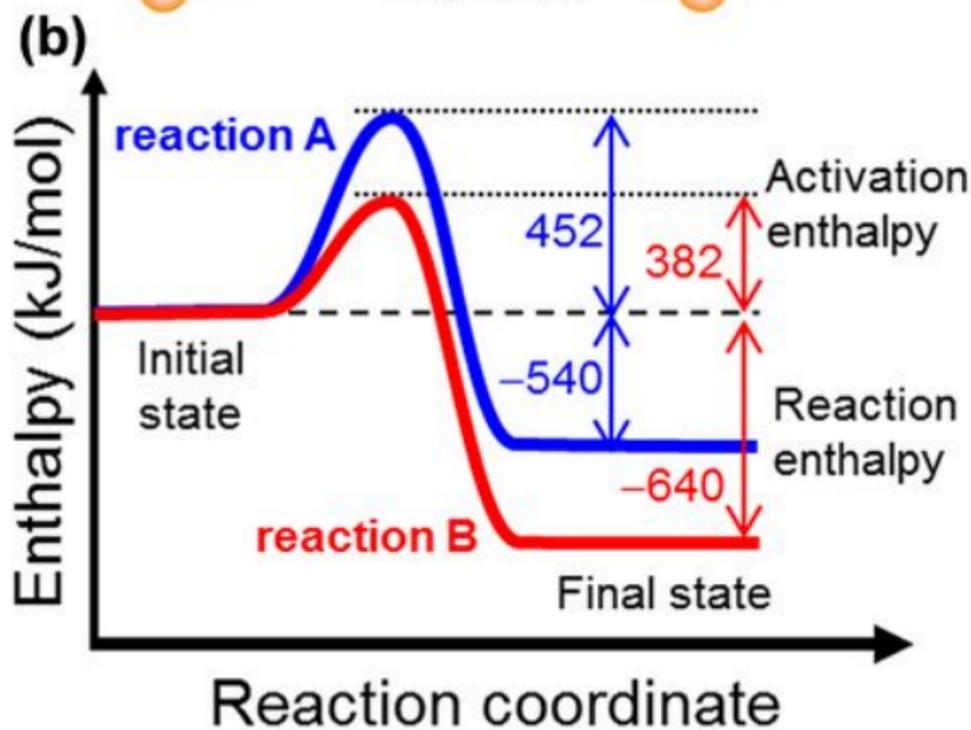

**Growth of LED structure**

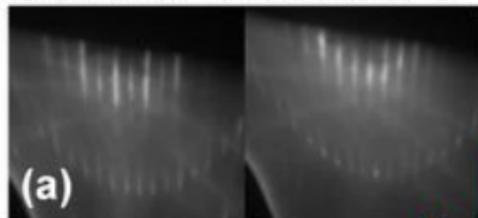
(a)

**1st Al epitaxial growth**

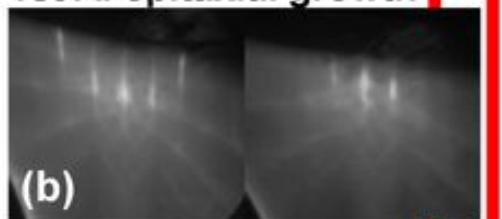
(b)

**1st oxidation**

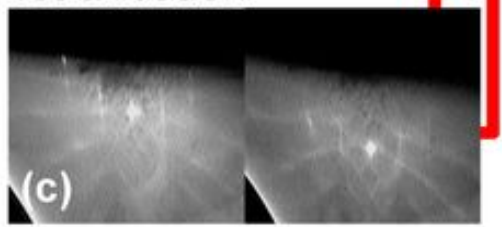
(c)

**2nd Al growth**

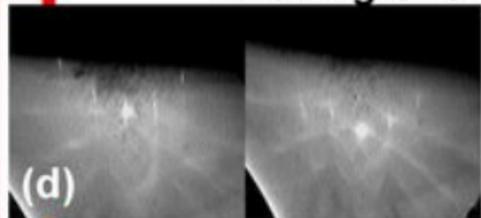
(d)

**2nd Oxidation**

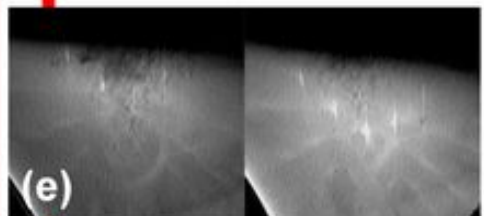
(e)

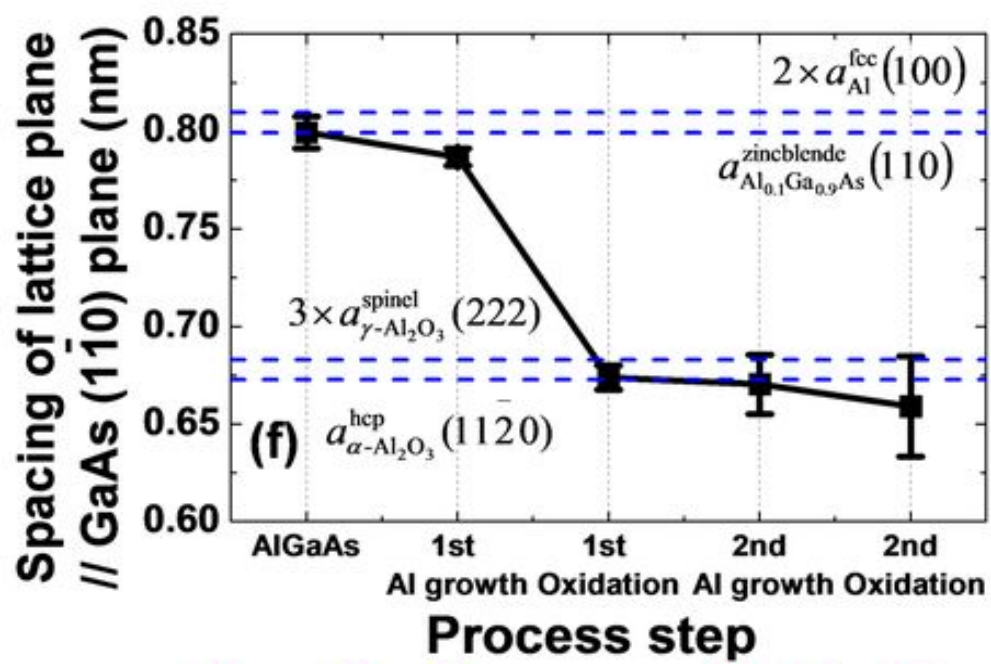

(f)

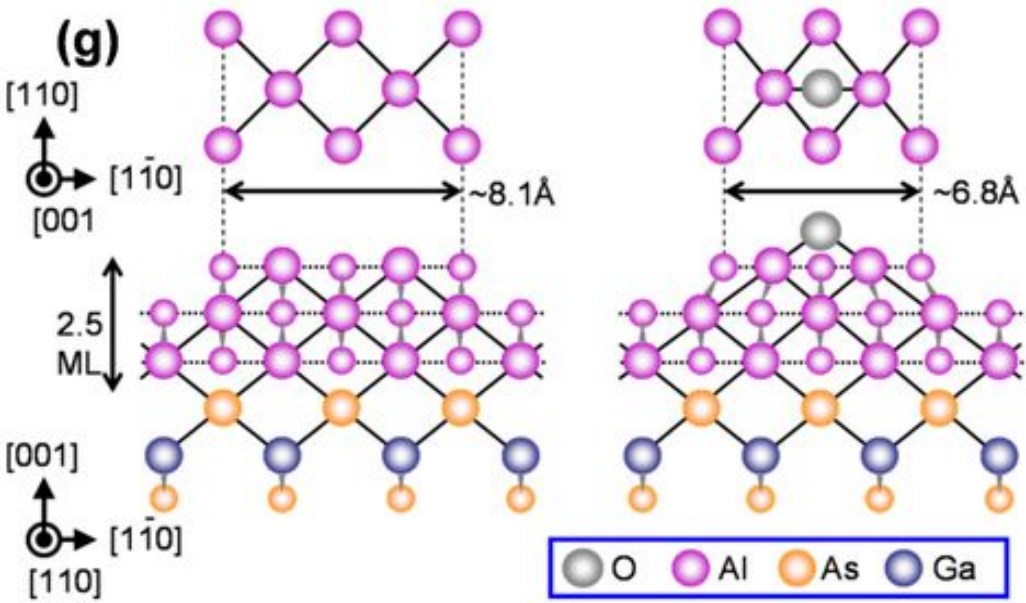

(g)

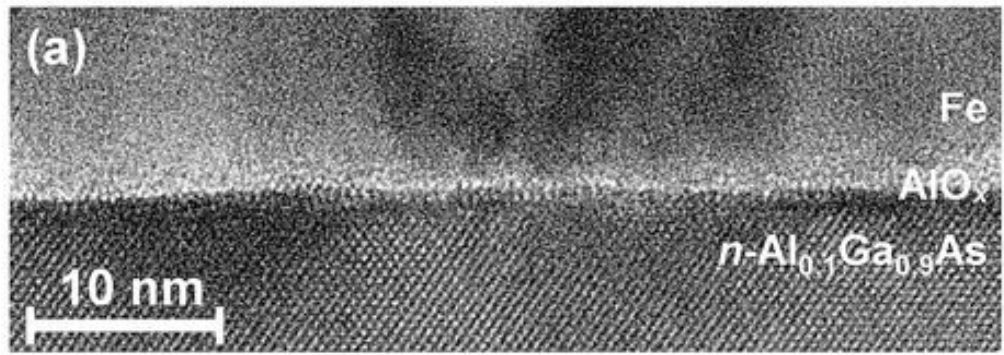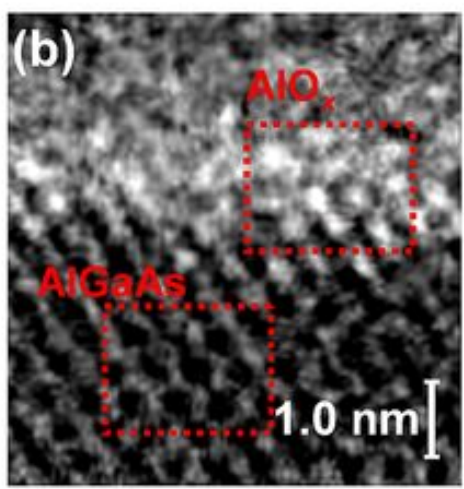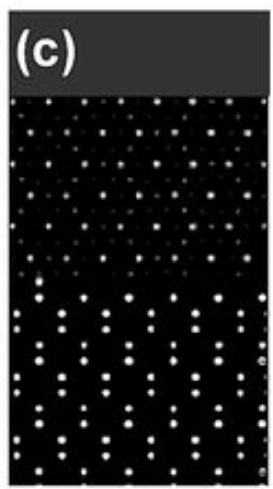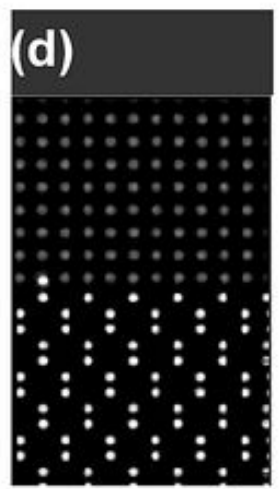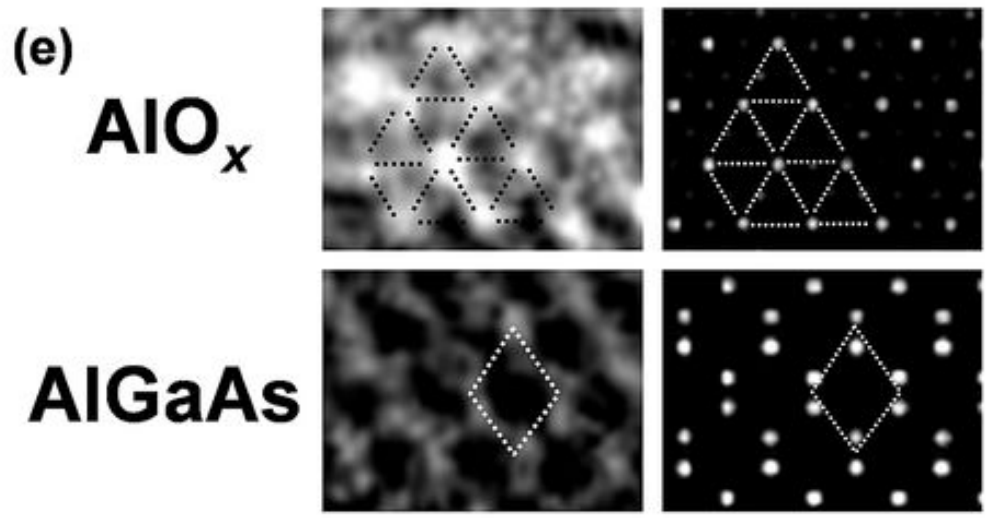

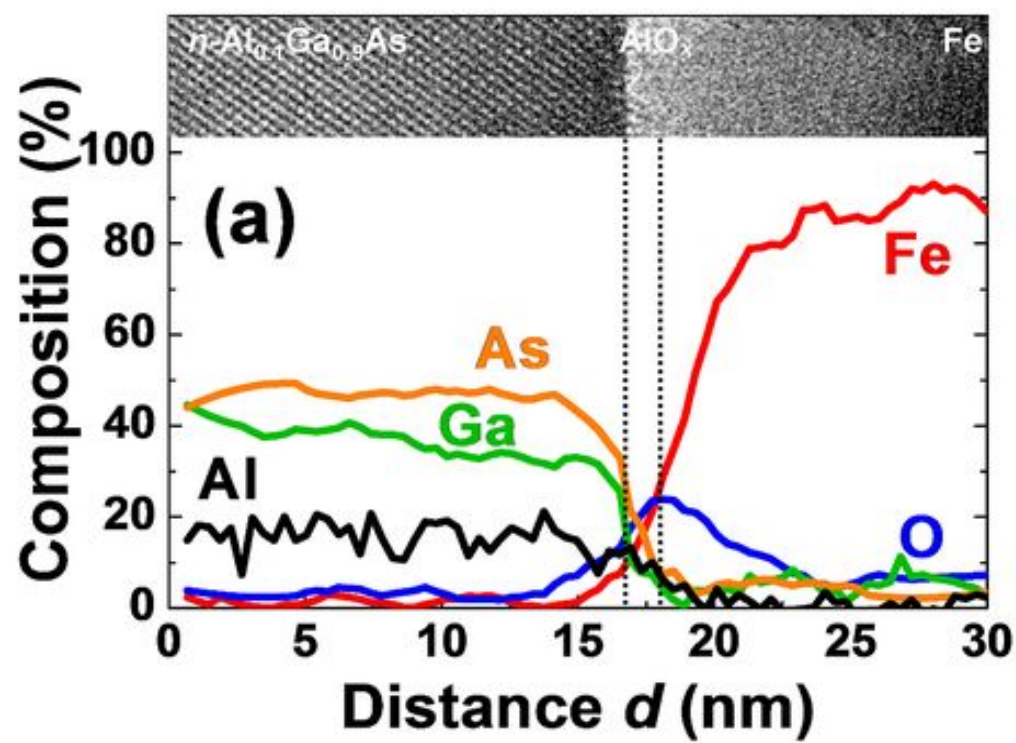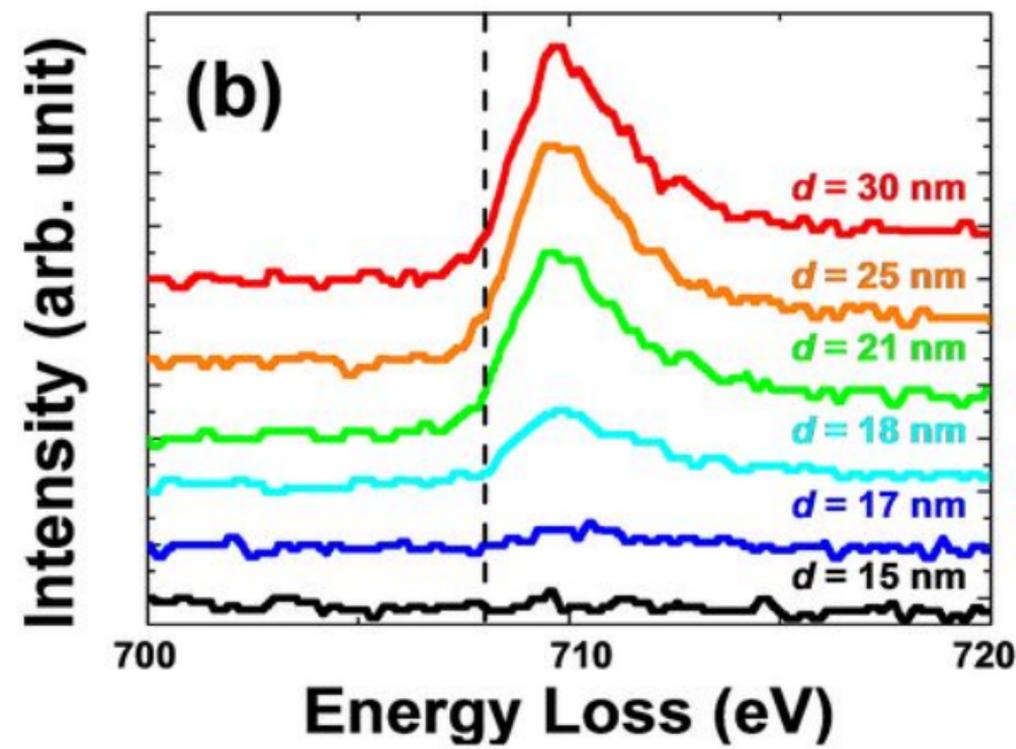

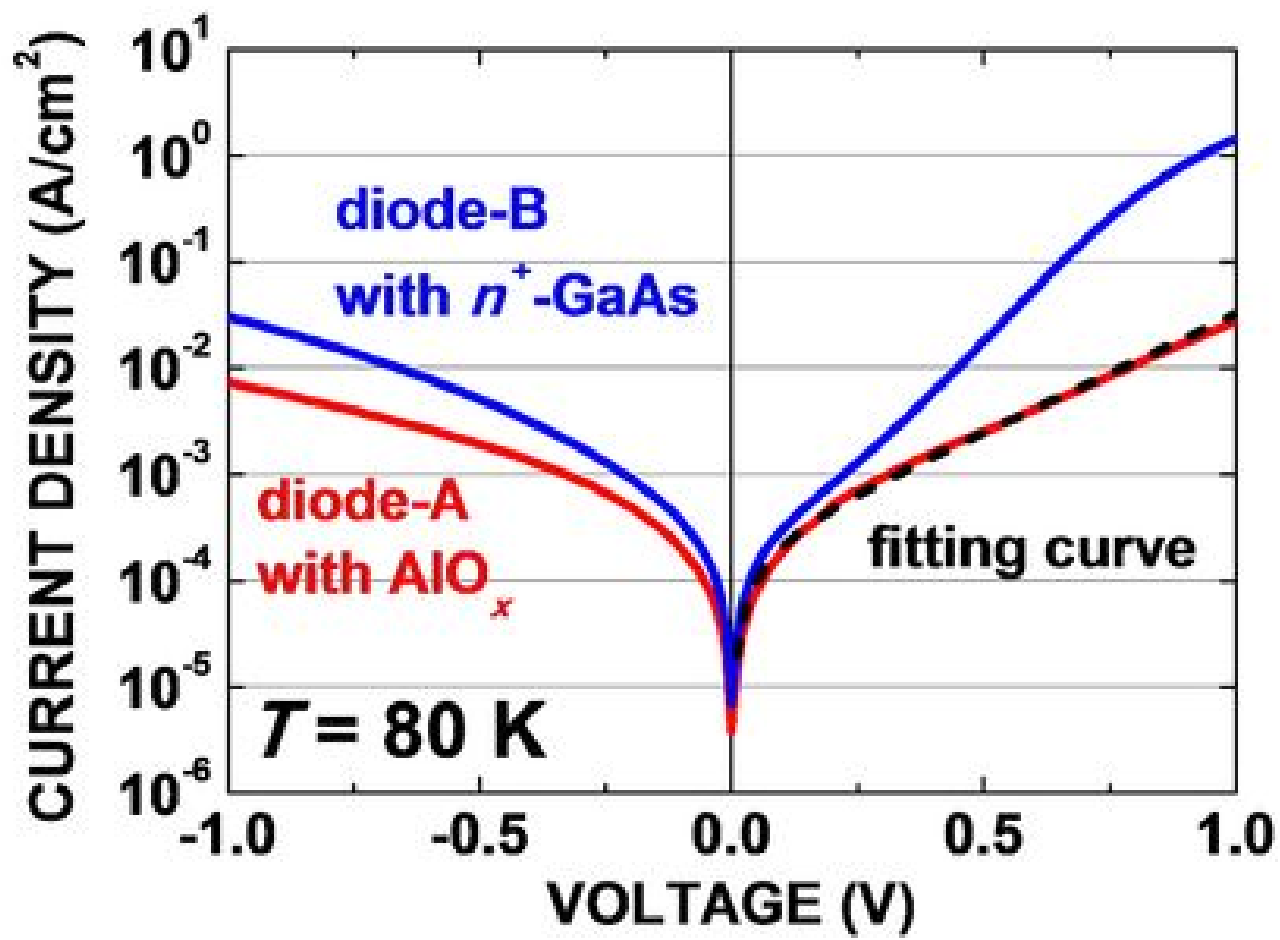

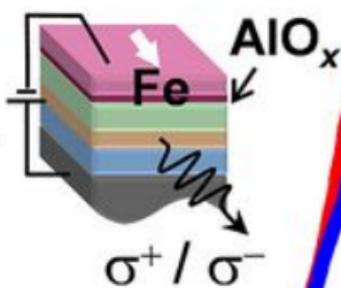
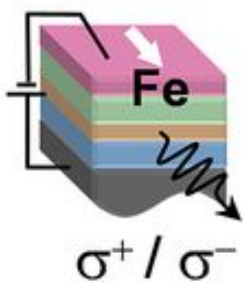